\documentclass[notoc]{ian}

\usepackage{amssymb}
\usepackage[utf8]{inputenc}
\usepackage[russian]{babel}
\usepackage{graphicx}

\begin{document}

\hbox{}
\vfill

\hfil{\bf\LARGE
Crystalline ordering and large fugacity expansion\par
\vskip10pt
\hfil for hard-core lattice particles
}
\vskip80pt

\hfil{\bf\large Ian Jauslin}\par
\hfil{\it School of Mathematics, Institute for Advanced Study}\par
\hfil{\tt\color{blue}\href{mailto:jauslin@ias.edu}{jauslin@ias.edu}}
\vskip20pt

\hfil{\bf\large Joel L. Lebowitz}\par
\hfil{\it Departments of Mathematics and Physics, Rutgers University}\par
\hfil{\it Simons Center for Systems Biology, Institute for Advanced Study}\par
\hfil {\tt\color{blue}\href{mailto:lebowitz@math.rutgers.edu}{lebowitz@math.rutgers.edu}}

\vskip80pt

\hfil {\bf Abstract}\par
\medskip
Using an extension of Pirogov-Sinai theory we prove phase transitions, corresponding to sublattice orderings, for a general class of hard-core lattice particle systems with a finite number of perfect coverings. These include many cases for which such transitions have been proven. The proof also shows that, for these systems, the Gaunt-Fisher expansion of the pressure in powers of the inverse fugacity (aside from an explicit logarithmic term) has a nonzero radius of convergence.

\vskip20pt

\hfil{\it Dedicated to Ben Widom, on the occasion of his 90th birthday.}

\vfill\eject

\setcounter{page}1
\pagestyle{plain}

\section{Introduction}
\indent The study of order-disorder phase transitions for hard-core lattice particle (HCLP) systems has a long history (see, to name but a few, \cite{Do58,Ka63,GF65,Ba80,Mc10,RD12,DG13} and references therein). These are purely entropy driven transitions, similar to those observed numerically and experimentally for hard spheres in the continuum~\-\cite{WJ57,AW57,PM86,IK15}. Whereas a proof of the transition in the hard sphere model is still lacking, there are several HCLP systems in which phase transitions have been proved.

\begin{figure}
  \hfil\includegraphics[width=2cm]{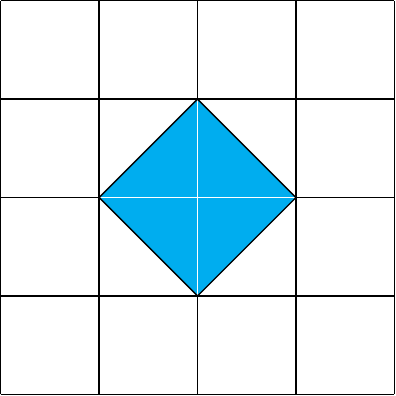}\ {\footnotesize\it a.}
  \hfil\includegraphics[width=2cm]{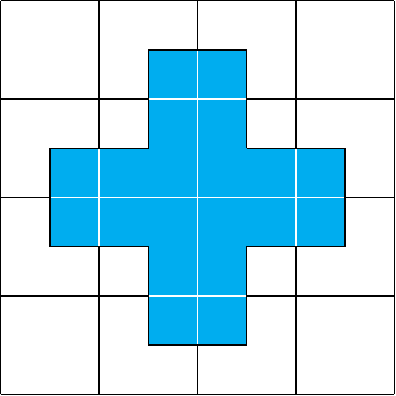}\ {\footnotesize\it b.}
  \medskip
  \caption{%
    {\it a}. A diamond on the square lattice. This system is equivalent to the nearest-neighbor exclusion model.\par
    {\it b}. A cross on the square lattice. This system is equivalent to the third-nearest-neighbor exclusion model.
  }
  \label{fig:shapes}
\end{figure}

\indent One example is the hard diamond model on the square lattice (see figure~\-\ref{fig:shapes}{\it a}), which is a particle model on $\mathbb Z^2$ with nearest-neighbor exclusion. The existence of a transition from a low-density disordered state, in which the density of occupied sites is the same on the even and odd sublattices, to a high-density ordered state, in which one of the sublattices is preferentially occupied, was proved by Dobrushin~\-\cite{Do68}, using a Peierls-type construction. This transition had been predicted earlier, using various approximations. In particular, Gaunt and Fisher~\-\cite{GF65} did an extensive study of this model using Pad\'e approximants, obtained from a low- and a high-fugacity expansion for the pressure $p(z)$, to determine the location of a singularity on the positive $z$ axis. They estimated that there is a transition at fugacity $z_t=3.8$ and density $\rho_t=0.37$, which is in good agreement with computer simulations.

\indent Another example is that of hard hexagons on a triangular lattice. Baxter~\-\cite{Ba80,Ba82} obtained, following earlier numerical and approximate work, an exact solution of this system, and found a transition at $z_t=\frac12(5\sqrt5+11)\approx11.09$ and $\rho_t=\frac1{10}(5-\sqrt5)\approx0.28$. There was further work, particularly by Joyce~\-\cite{Jo88}, which elaborated on this solution. Baxter's solution provides a full, albeit implicit, expression for the pressure $p(z)$ in the complex $z$ and $\rho$ planes, implying, in particular, that $p(z)$ is analytic for all $z\geqslant 0$ except at $z_t$. The transition at $z_t$ is of second order, as is expected to be the case for diamonds. For $z>z_t$, this system has 3 ordered phases, corresponding to the 3 different perfect coverings.

\begin{figure}
  \hfil\includegraphics[height=6cm]{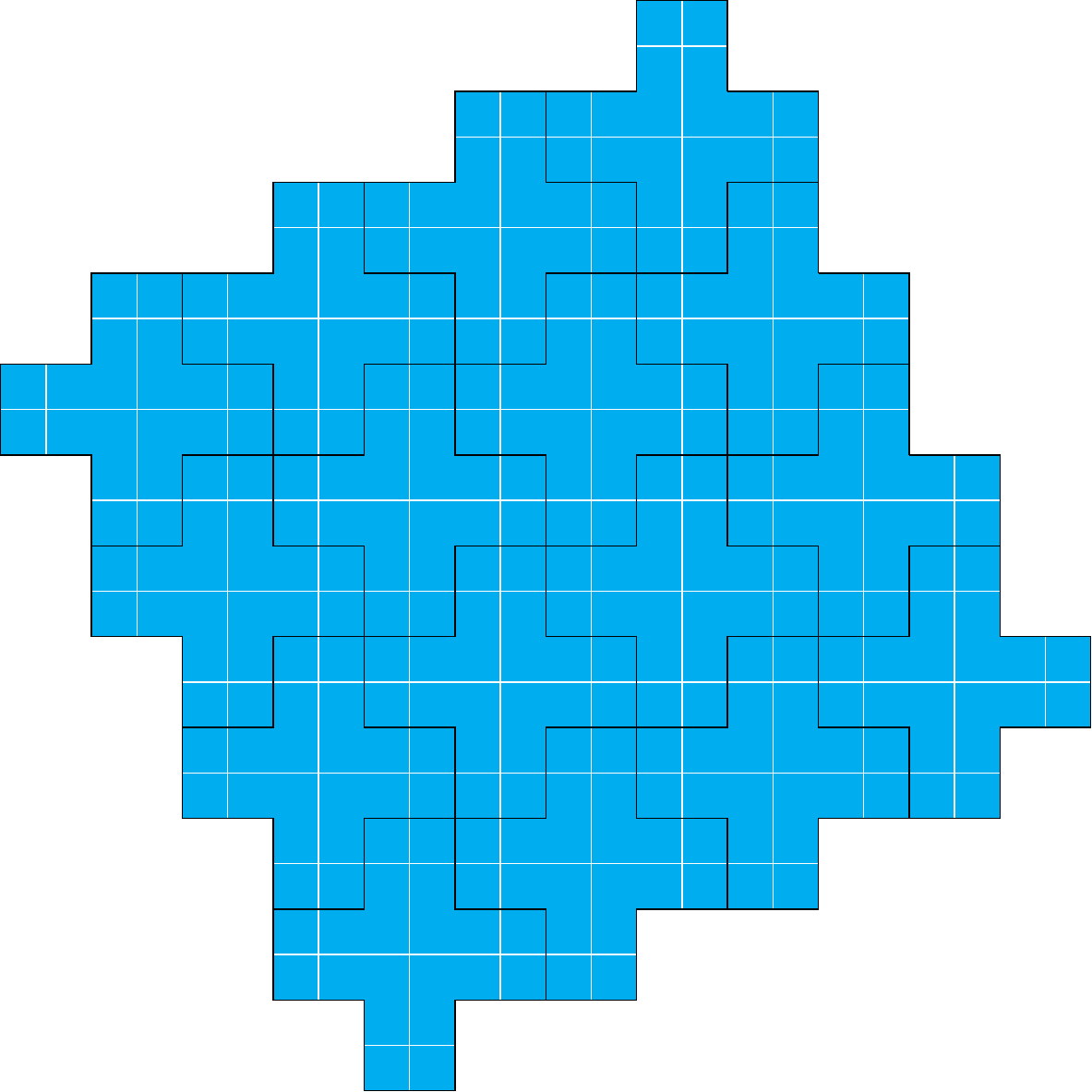}
  \hfil\includegraphics[height=6cm]{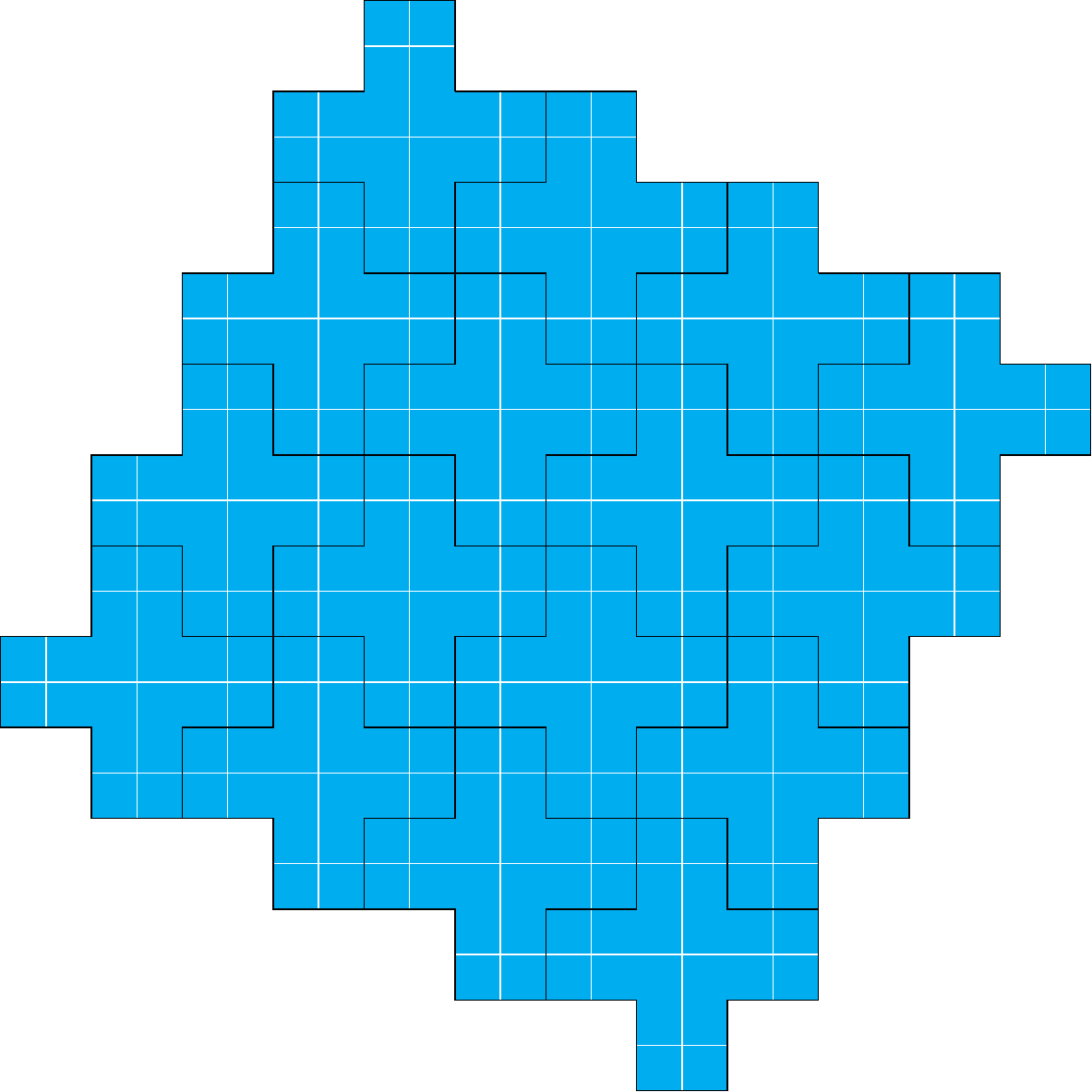}
  \medskip
  \caption{The perfect coverings of the cross model are obtained by translating these two configurations. For each figure, there are 5 inequivalent translations, thus totalling 10 perfect coverings.}
  \label{fig:cross_packing}
\end{figure}

\indent Yet another HCLP model for which an order-disorder transition was shown to occur, with only a sketch of a proof~\-\cite{HP74}, is that of hard crosses on the square lattice (see figure~\-\ref{fig:shapes}{\it b}). This model has 10 distinct perfect coverings (see figure~\-\ref{fig:cross_packing}), and is conjectured~\-\cite{EB05} to have a first order phase transition at $z_t\approx39.5$. At this fugacity the density jumps from $\rho_f\approx0.16$ to $\rho_s\approx0.19$. We shall use this model as an illustration for the type of system to which our analysis applies, and for which we can prove crystalline order at high fugacities, and the convergence of the high-fugacity expansion.
\bigskip

\indent The expansion in powers of $y\equiv z^{-1}$ was first considered by Gaunt and Fisher~\-\cite{GF65} specifically for the diamond model, but has been used later for other HCLP systems~\-\cite{Jo88,EB05}. As far as we know, there has been no study of the convergence of this series, though Baxter's explicit solution~\-\cite{Ba80} for the hard hexagon model shows it is so for that solvable model. This is in contrast to the low-fugacity expansion of the pressure $p(z)$ in powers of $z$, which dates back to Ursell~\-\cite{Ur27} and Mayer~\-\cite{Ma37}. This expansion was proven to have a positive radius of convergence, in all dimensions, by Groeneveld~\-\cite{Gr62} for positive pair-potentials and by Ruelle~\-\cite{Ru63} and Penrose~\-\cite{Pe63} for general pair-potentials.

\bigskip

\indent In this note, we sketch a proof that the radius of convergence of the high-fugacity expansion is positive for a certain class of HCLP systems in $d\geqslant 2$ dimensions. For details of the proof, see~\-\cite{JL17}. The proof is based on an extension of Pirogov-Sinai theory~\-\cite{PS75,KP84}, and implies the existence of phase transitions in these models. Unlike the low-fugacity expansion, the positivity of the radius of convergence does not hold for general HCLP systems: there are, indeed, many examples, in 1 and higher dimensions, in which the coefficients in this expansion diverge in the thermodynamic limit.
\vskip20pt

\subsection{Description of the model}\label{sec:model}
\indent It is convenient, for our analysis, to think of these HCLPs as having a finite shape $\omega$ in physical space $\mathbb R^d$, and impose the constraint that, when put on lattice sites $x$ and $y$, the shapes do not overlap (for example, for the nearest-neighbor exclusion on the square lattice, $\omega$ could be a diamond, see figure~\-\ref{fig:shapes}{\it a}). Equivalently, one can think of each particle as occupying a finite collection of lattice sites. Note that the choice of these shapes, or of the lattice points assigned to each particle, is not unique: two different shapes can translate to the same hard-core interaction. For instance, the nearest neighbor exclusion on the square lattice can be obtained by taking diamonds or disks of radius $r$ with $\frac12<r<\frac1{\sqrt2}$.
\bigskip

\indent Given a $d$-dimensional lattice $\Lambda_\infty$ and a finite subset $\Lambda\subset\Lambda_\infty$, we define the grand-canonical partition function of the system at activity $z>0$ on $\Lambda$, with some specified boundary conditions, as
\begin{equation}
  \Xi_\Lambda(z)=\sum_{X\subset\Lambda}z^{|X|}\prod_{x\neq x'\in X}\varphi(x,x')
  \label{Xi}
\end{equation}
in which $X$ is a particle configuration in $\Lambda$, $|X|$ is the number of particles, and, setting $\omega_x\equiv \{x+y,\ y\in\omega\}$, $\varphi(x,x')\in\{0,1\}$ enforces the hard-core repulsion: it is equal to 1 if and only if $\omega_{x}\cap\omega_{x'}=\emptyset$. Note that, due to the hard-core interaction, the number of particles is bounded:
\begin{equation}
  |X|\leqslant N_{\mathrm{max}}\leqslant|\Lambda|
\end{equation}
where $|\Lambda|$ denotes the number of lattice sites in $\Lambda$. Our aim, in this note, is to prove that, in certain cases, the finite-volume {\it pressure} of the system, defined as
\begin{equation}
  p_\Lambda(z):=\frac1{|\Lambda|}\log \Xi_\Lambda(z)
  \label{p}
\end{equation}
satisfies
\begin{equation}
  p(z):=\lim_{\Lambda\to\Lambda_\infty}p_\Lambda=\rho_m\log z+f(y)
\end{equation}
in which $\rho_m$ is the maximum density in $\Lambda$, that is, $\rho_m=\lim_{\Lambda\to\Lambda_\infty}N_{\mathrm{max}}/|\Lambda|$, and $f$ is an analytic function of $y\equiv z^{-1}$ for small values of $y$. The expansion of $f$ in powers of $y$ is called the {\it high-fugacity expansion} of the system. Note that, as is well known, $p(z)\equiv\lim_{\Lambda\to\Lambda_\infty}p_\Lambda(z)$ is independent of the boundary conditions for all $z\geqslant 0$ (see, for instance, \cite{Ru99}). This is not so for the correlation functions, which may depend on the boundary conditions, and can also be shown to be analytic in $y$ with the same radius of convergence as the high-fugacity expansion.

\subsection{Low-fugacity expansion}\label{sec:low_fugacity}
\indent It is rather straightforward to express the pressure $p_\Lambda$ as a power series in $z$ (which converges for small values of $z$, thus earning the name ``low-fugacity expansion''). Indeed, defining the {\it canonical} partition function by $Z_\Lambda(k)$ as the number of particle configurations with $k$ particles, (\ref{Xi}) can be rewritten as
\begin{equation}
  \Xi_\Lambda(z)=\sum_{k=0}^{N_{\mathrm{max}}} z^kZ_\Lambda(k).
  \label{Xi_z}
\end{equation}
Substituting~\-(\ref{p}) into~\-(\ref{Xi_z}), we find that, formally,
\begin{equation}
  p_\Lambda=\sum_{k=1}^\infty z^kb_k(\Lambda)
  \label{p_z}
\end{equation}
with
\begin{equation}
  b_k(\Lambda):=\frac1{|\Lambda|}\sum_{n=1}^k\frac{(-1)^{n+1}}n\sum_{\displaystyle\mathop{\scriptstyle k_1,\cdots,k_n\geqslant 1}_{k_1+\cdots+k_n=k}}Z_\Lambda(k_1)\cdots Z_\Lambda(k_n).
  \label{blog}
\end{equation}
As is well known (see, for instance, \cite{Ru99}), there is a remarkable cancellation that eliminates the $\Lambda$ dependence from $b_k(\Lambda)$ when $\Lambda\to\Lambda_\infty$. This is readily seen by writing these coefficients in terms of {\it Mayer graphs}, which implies that $b_j(\Lambda)$ converges to $b_j$ as $\Lambda\to\Lambda_\infty$, independently of the boundary condition. Furthermore, the radius of convergence $R(\Lambda)$ of~\-(\ref{p_z}) converges to $R>0$, which is equal, for positive potentials (like those considered here), to the radius of convergence $R_\infty$ of $\sum_{j=1}^\infty b_jz^j$ \cite{Pe63}.

\subsection{High-fugacity expansion}
\indent The main idea of the high-fugacity expansion, due to Gaunt and Fisher~\-\cite{GF65}, is to perform a {\it low-fugacity} expansion for the {\it holes} of the system. In other words, instead of expressing the pressure $p_\Lambda$ in terms of the number of {\it particle} configurations with $k$ particles, we express it in terms of the number of {\it hole} configurations in the {\it absence} of $k$ particles from perfect covering. To make this idea more precise, let us consider the example of nearest-neighbor exclusion (which corresponds to hard diamonds) on the square lattice.
\bigskip

\indent Assume that $\Lambda$ is a $2n\times 2n$ torus, so that $\Lambda$ can be completely packed with diamonds. Note that, by the invariance of the system under translations, there are two perfect coverings, each of which contains $|\Lambda|/2$ particles. Let
\begin{equation}
  Q_\Lambda(k):=Z_\Lambda(\textstyle\frac{|\Lambda|}2-k)
\end{equation}
denote the number of configurations that are {\it missing} $k$ particles, in terms of which
\begin{equation}
  \Xi_\Lambda(z)=2z^{\frac12|\Lambda|}\sum_{k=0}^\infty \left(\frac12z^{-k}Q_\Lambda(k)\right)
  \label{Xi_hole}
\end{equation}
(we factor the $2$ out, because $Q_\Lambda(0)=2$ and we wish to expand the logarithm in~\-(\ref{p}) around 1). We thus have, formally
\begin{equation}
  p_\Lambda=\frac1{|\Lambda|}\log2+\frac12\log z+\sum_{k=1}^\infty y^{k}c_k(\Lambda)
  \label{p_y}
\end{equation}
where $y\equiv z^{-1}$ and
\begin{equation}
  c_k(\Lambda):=\frac1{|\Lambda|}\sum_{n=1}^k\frac{(-1)^{n+1}}{n2^n}\sum_{\displaystyle\mathop{\scriptstyle k_1,\cdots,k_n\geqslant 1}_{k_1+\cdots+k_n=k}}Q_\Lambda(k_1)\cdots Q_\Lambda(k_n).
  \label{clog}
\end{equation}
The first nine terms of this expansion were computed in~\-\cite[table~\-XIII]{GF65} for periodic boundary conditions. They found that, as in the low-fugacity expansion, these nine coefficients $c_k(\Lambda)$ converge to $c_k$ as $\Lambda\to\Lambda_\infty$. However, there is no systematic way of exhibiting the cancellations needed for this convergence to hold for general HCLP systems. In fact there are many simple examples where $c_2(\Lambda)\to\infty$ as $\Lambda\to\Lambda_\infty$. For instance, consider the nearest-neighbor exclusion in 1 dimension (which maps, exactly, to the one-dimensional monomer-dimer model). In the language of this paper, the model is characterized by $\Lambda_\infty=\mathbb Z$, $\omega=(-1,1)\subset\mathbb R$. It is easy to compute, using, for instance, the transfer matrix technique, that the infinite-volume pressure of this model is
\begin{equation}
  p=\log\left(\frac{1+\sqrt{1+4z}}2\right)
  =\frac12\log z+\log\left(\sqrt{1+\frac1{4z}}+\frac1{2\sqrt z}\right).
\end{equation}
Therefore, $p-\frac12\log z$ (note that $\rho_m=\frac12$) is not an analytic function of $y\equiv z^{-1}$ at $y=0$ (though it is an analytic function of $\sqrt y$). For the $n$-nearest-neighbor exclusion in one dimension, $p-\rho_m\log z$ is analytic in $y^{\frac1n}$. Similar effects occur in higher dimensions as well, for instance in systems exhibiting columnar order at high fugacities~\-\cite{GD07}.
\bigskip

\indent Note that, for systems whose pressure admits a convergent high-fugacity expansion, the partition function may not have any roots for large values of $|z|$, which implies that the Lee-Yang~\-\cite{YL52,LY52} zeros of such systems are confined within an annulus: denoting the radius of convergence of the low- and high-fugacity expansions by $R$ and $\tilde R$, every Lee-Yang zero $\xi$ satisfies
\begin{equation}
  R\leqslant|\xi|\leqslant \tilde R^{-1}
  .
\end{equation}
Furthermore, it can easily be seen (by Krammers-Wannier duality) that systems with bounded repulsive pair-potentials all have a convergent high-fugacity expansion, so their Lee-Yag zeros lie within an annulus.
\bigskip

\indent Here, we prove that, for a class of HCLP systems which we call ``non-sliding models'' (which include the three models discussed above, that is, the hard diamond, cross and hexagon models), the function
\begin{equation}
  f_\Lambda(y):=p_\Lambda-\rho_m\log z+o(1)
  \label{f}
\end{equation}
is analytic in a disk around $y=0$ uniformly in $|\Lambda|$, in which $o(1)\to0$ as $\Lambda\to\Lambda_\infty$. That is,
\begin{equation}
  f_\Lambda(y)=\sum_{k=1}^\infty y^kc_k(\Lambda)
  ,\quad
  |c_k(\Lambda)|<\tilde R^k
  ,\quad
  \lim_{\Lambda\to\Lambda_\infty}c_k(\Lambda)=c_k
  \label{goal}
\end{equation}
for some $\tilde R>0$, independent of $|\Lambda|$. We thus prove the validity of the Gaunt-Fisher expansion for non-sliding models. Our method of proof further shows that, for such systems, the high-fugacity phases exhibit crystalline order.
\bigskip

\begin{figure}
  \hfil\includegraphics[width=4cm]{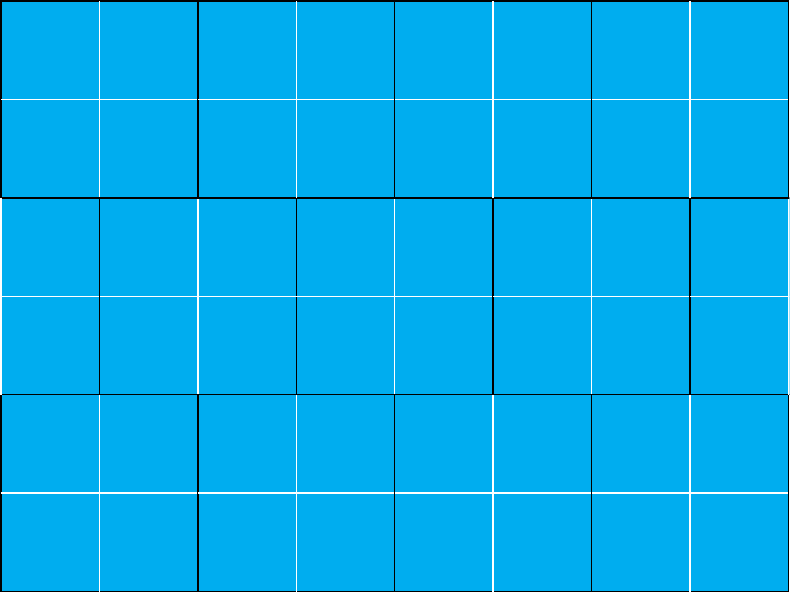}
  \caption{Hard $2\times2$ squares on the square lattice slide: whole columns or rows of particles can be moved without forming vacancies.}
  \label{fig:sliding}
\end{figure}

\indent A precise definition of the notion of non-sliding will be given below. An example of a {\it sliding} model is the hard $2\times2$ square model on the square lattice: given a perfect covering, whole columns or rows of particles can slide without forming vacancies (see figure~\-\ref{fig:sliding}). On the other hand, hard diamonds do {\it not slide}: the close-packed configurations are rigid. The same is true of hard crosses and hard hexagons, as well as the nearest neighbor exclusion on $\mathbb Z^d$ for any $d\geqslant 2$.

\section{Sketch of the proof}

\indent In this note, we will only give a detailed sketch of the proof. The full details can be found in~\-\cite{JL17}.
\bigskip

\indent Ultimately, the reasoning behind~\-(\ref{goal}) is similar to that underpinning the convergence of the Mayer expansion~\-(\ref{p_z}), so let us first discuss the Mayer expansion, and, in particular, focus on the uniform boundedness of $b_k(\Lambda)$ defined in~\-(\ref{blog}). For the sake of simplicity, we will consider periodic boundary conditions, and take $|\Lambda|$ sufficiently larger than $k$. First of all, note that $Z_\Lambda(k)$ is a polynomial in $|\Lambda|$ of order $k$ with no constant term, thus,
\begin{equation}
  \frac1{|\Lambda|}Z_\Lambda(k_1)\cdots Z_\Lambda(k_n)
\end{equation}
is a polynomial in $|\Lambda|$ of order $k-1$. Therefore, in order for $b_k(\Lambda)$ to remain bounded in the $\Lambda\to\Lambda_\infty$ limit for $k>1$, there must be a significant cancellation. For the purpose of illustration, consider
\begin{equation}
  b_2(\Lambda)=\frac1{|\Lambda|}\left(Z_\Lambda(2)-\frac12Z_\Lambda(1)^2\right).
\end{equation}
In the 1-particle case, the particle can occupy any site in $\Lambda$, so $Z_\Lambda(1)=|\Lambda|$. In the 2-particle case, the particles must not overlap. We can therefore write $Z_\Lambda(2)$ as the number of unconstrained configurations (excluding the cases in which particles coincide) minus the number of configurations in which the particles overlap. The former is equal to $\frac12|\Lambda|(|\Lambda|-1)$, and the latter is proportional to $|\Lambda|$. The $|\Lambda|^2$ term thus cancels out. This reasoning can be extended to all $b_k(\Lambda)$.
\bigskip

\begin{figure}
  \hfil
  \includegraphics[height=6cm]{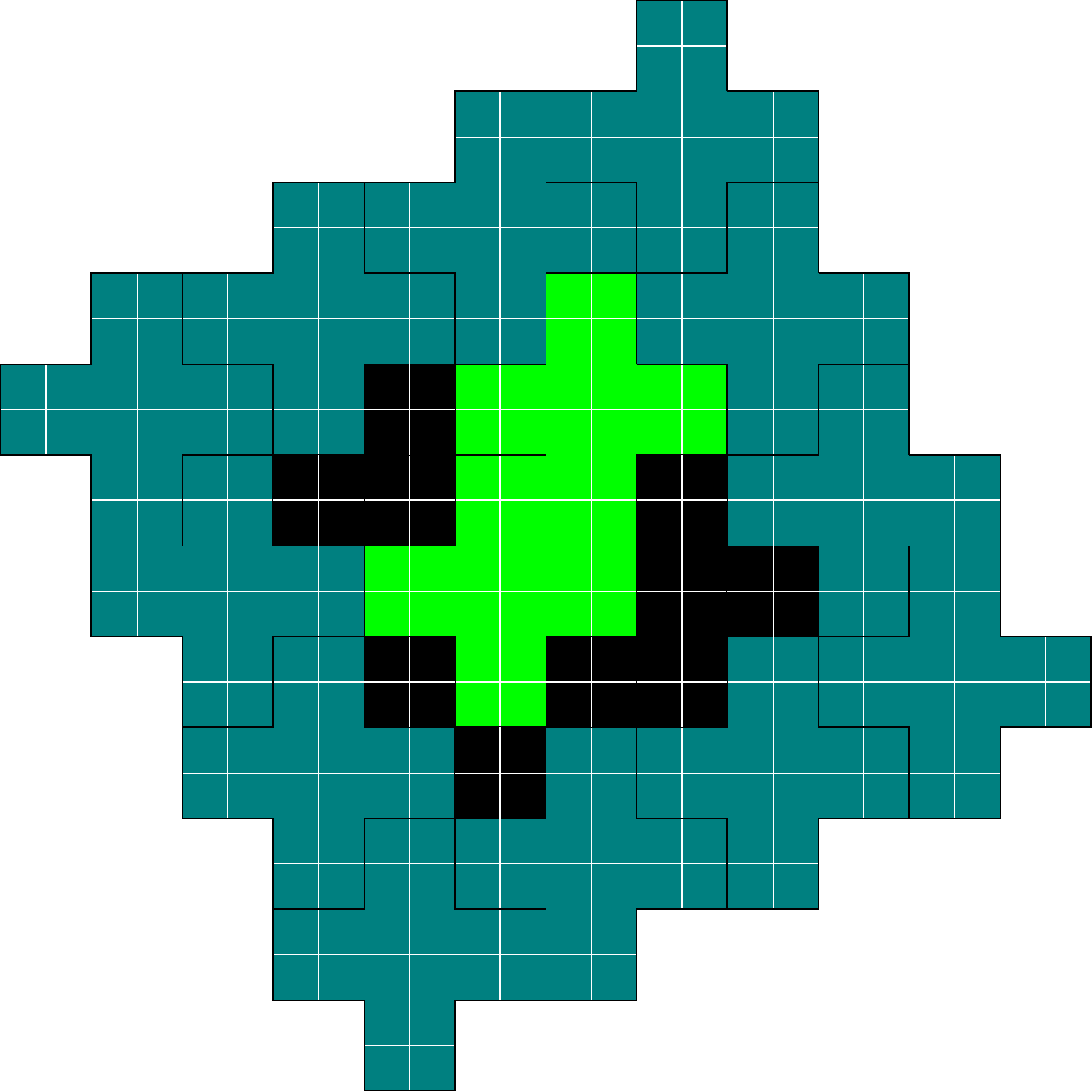}\hskip-20pt{\footnotesize\it a.}
  \hfil
  \includegraphics[height=6cm]{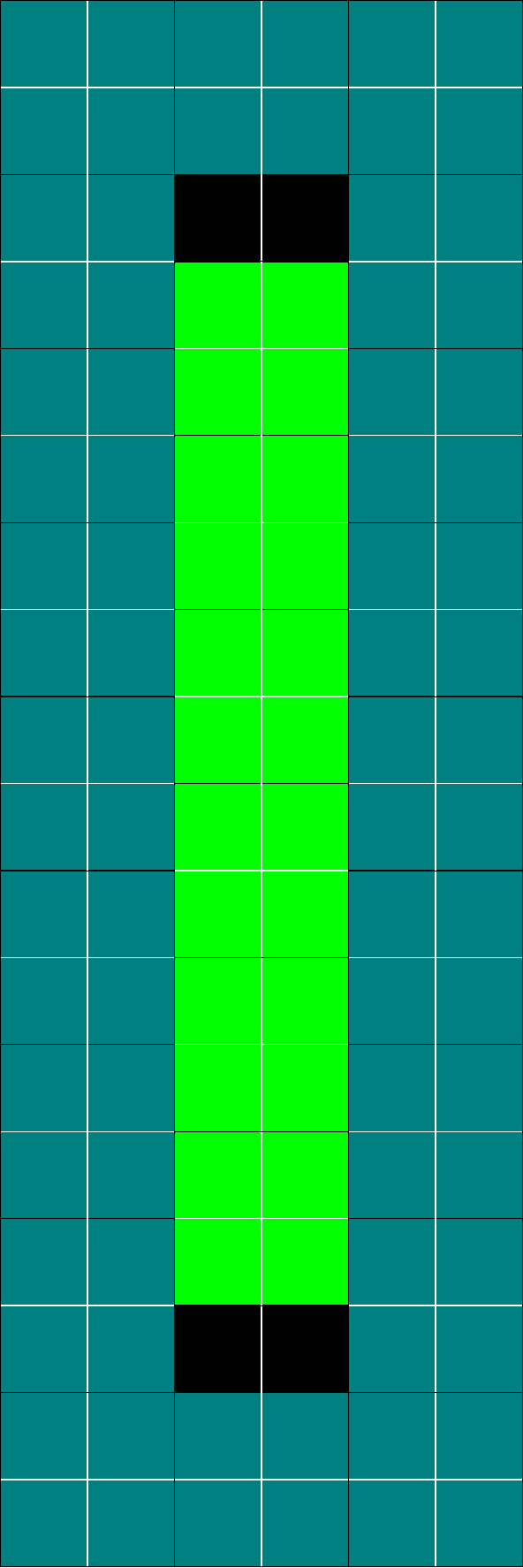}\ {\footnotesize\it b.}
  \medskip
  \caption{%
    {\it a}. A hole configuration for a system of hard crosses. There is no sliding in this model.
    \par
    {\it b}. A hole configuration for a system of hard $2\times2$ squares on the square lattice (next-nearest-neighbor exclusion). There is sliding in this model. Note that the central column corresponds to a configuration in the 1-dimensional nearest-neighbor exclusion model, which, as discussed earlier does not have an expansion in $z^{-1}$.
    \par
    \smallskip
    Particles of different colors correspond to different phases: the colored sub-configuration can be extended to different perfect coverings (note that the color assignment is not unique).
  }
  \label{fig:hole_example}
\end{figure}

\indent Following~\-\cite{GF65}, we construct the high-fugacity expansion in a similar way, but instead of counting particle configurations, we count hole configurations. To that end, we factor out $z^{\rho_m|\Lambda|}$ from the partition function, as in~\-(\ref{Xi_hole}), thus giving each hole a weight $z^{-\rho_m}$. The most significant difference with the Mayer expansion is that the interaction between holes is not simply a hard-core repulsion, since the sites that are not empty must be covered by particles which, in turn, must satisfy the hard-core constraint. In particular, the connected components of the empty space may come in various shapes and sizes, but they are constrained by the fact that the overall empty volume is an integer multiple of the volume of each particle (see figure~\-\ref{fig:hole_example}). This implies that different connected components of the empty volume could, in principle, interact strongly, even if they are arbitrarily far from each other (see figure~\-\ref{fig:hole_example}{\it b}). If that were the case, then the $|\Lambda|^2$ terms in $c_2(\Lambda)$ would not cancel out, and the high-fugacity expansion would be ill-defined in the thermodynamic limit. This phenomenon will be called {\it sliding}.
\bigskip

\begin{figure}
  \hfil\includegraphics[width=9cm]{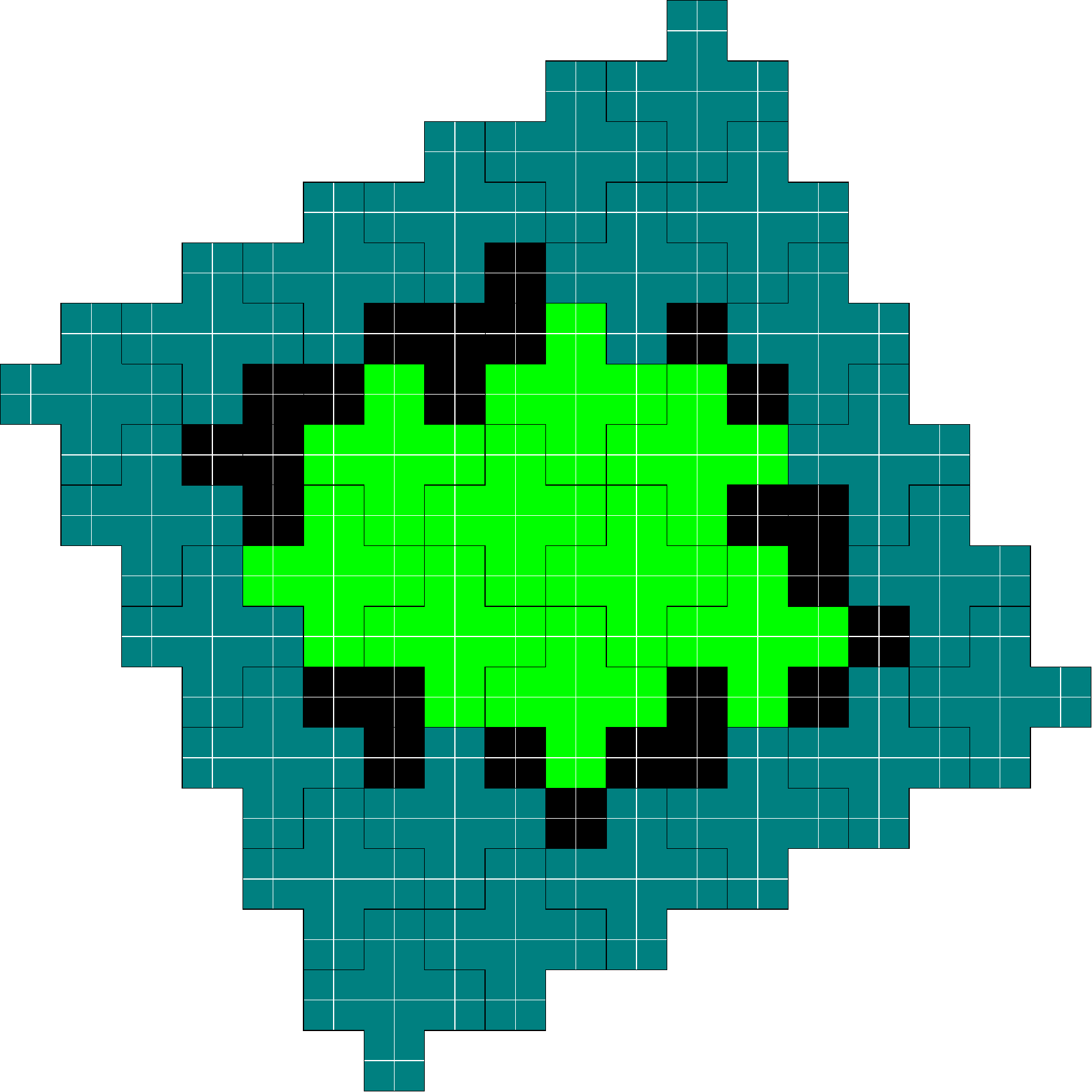}
  \caption{An interface between two different phases. The particles are colored according to the phase to which they belong (the coloring is not, in general, unique). There is empty space, in black in the figure, along the entire interface. This follows from the absence of sliding: for every connected collection of particles along the interface, there is empty space nearby.}
  \label{fig:interface}
\end{figure}

\indent In this paper, we will only consider models in which there is {\it no sliding}, a notion which we will now define precisely. First of all, in order to qualify as a non-sliding model, the system must only admit a {\it finite} number of distinct perfect coverings, and must be such that any particle configuration is entirely determined by the location of the holes and the particles adjacent to them. In addition, whenever different locally close-packed phases coexist, the interface between the phases must contain a number of holes proportional to its length (see figure~\-\ref{fig:interface}). This condition is analogous to the {\it Peierls condition} in the standard Pirogov-Sinai theory~\-\cite{PS75,KP84}. This rules out situations similar to figure~\-\ref{fig:hole_example}{\it b}, in which the interface between the central column and the other two may be arbitrarily long, while having only two holes. More precisely, a model is said to exhibit {\it no sliding} if, for every {\it connected} particle configuration $X\subset\Lambda$ that  is {\it not} a subset of a perfect covering of $\Lambda$, and for every configuration $Y\supset X$, there exists at least one empty site {\it neighboring} a particle in $X$ (see figure~\-\ref{fig:interface}). Note that, having fixed $X$, there are many possible connected configurations $Y$ that contain $X$, and we require that {\it every one} of them contain some empty space. The notions of {\it connectedness} and {\it neighbors} are inherited from the lattice structure, and a particle configuration $X$ is said to be connected if the set of lattice sites that are covered by particles is connected.
\bigskip

\begin{figure}
  \hfil\includegraphics[height=3.5cm]{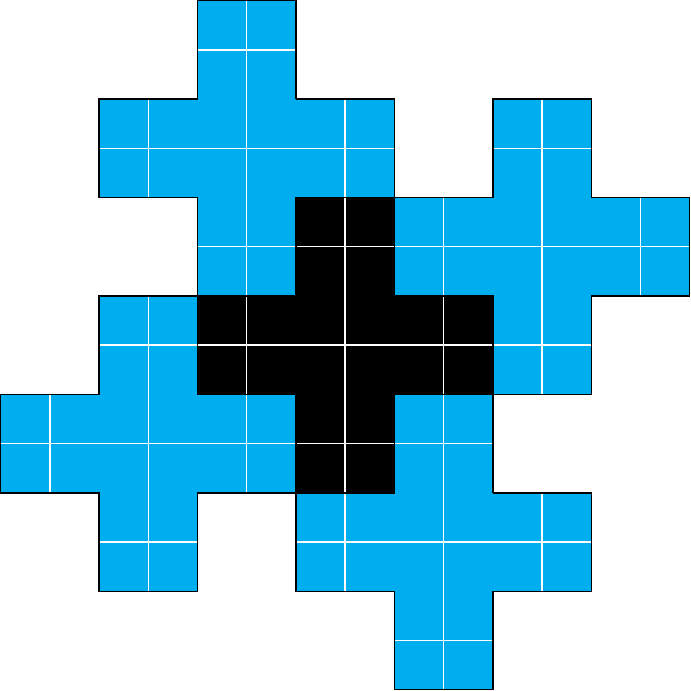}{\footnotesize\it a}.
  \hfil\includegraphics[height=4.5cm]{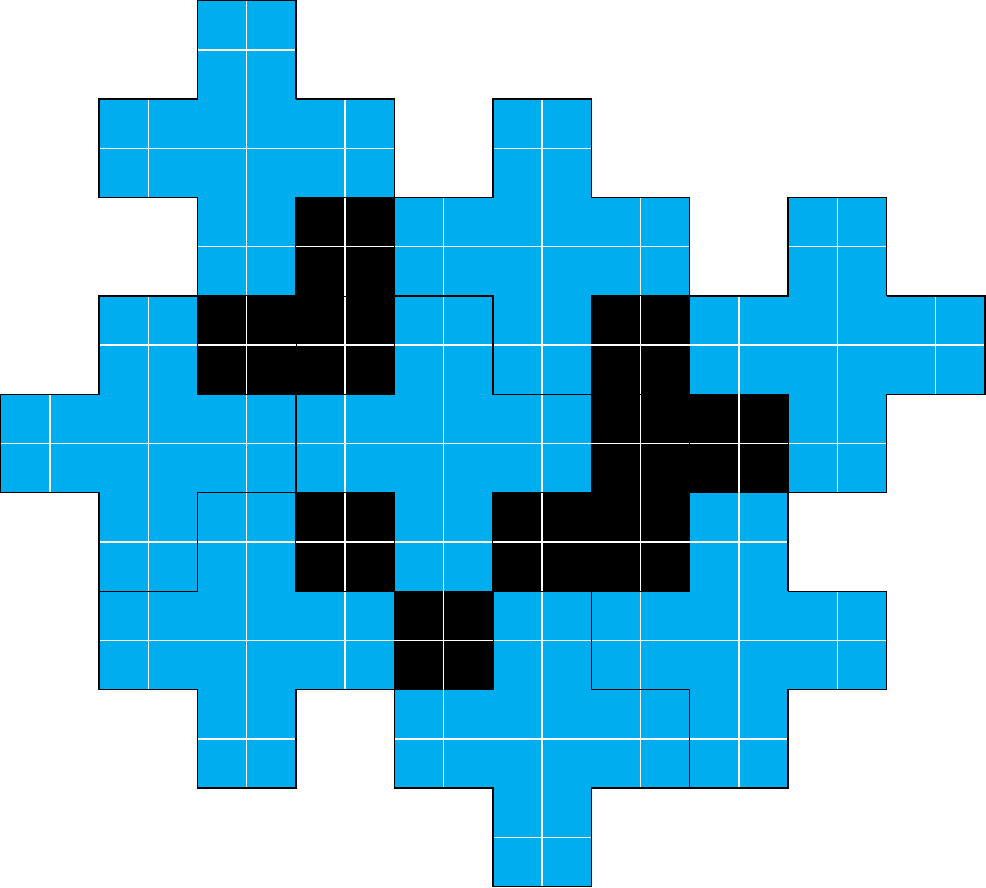}{\footnotesize\it b}.\par\penalty10000
  \bigskip
  \hfil\includegraphics[height=7cm]{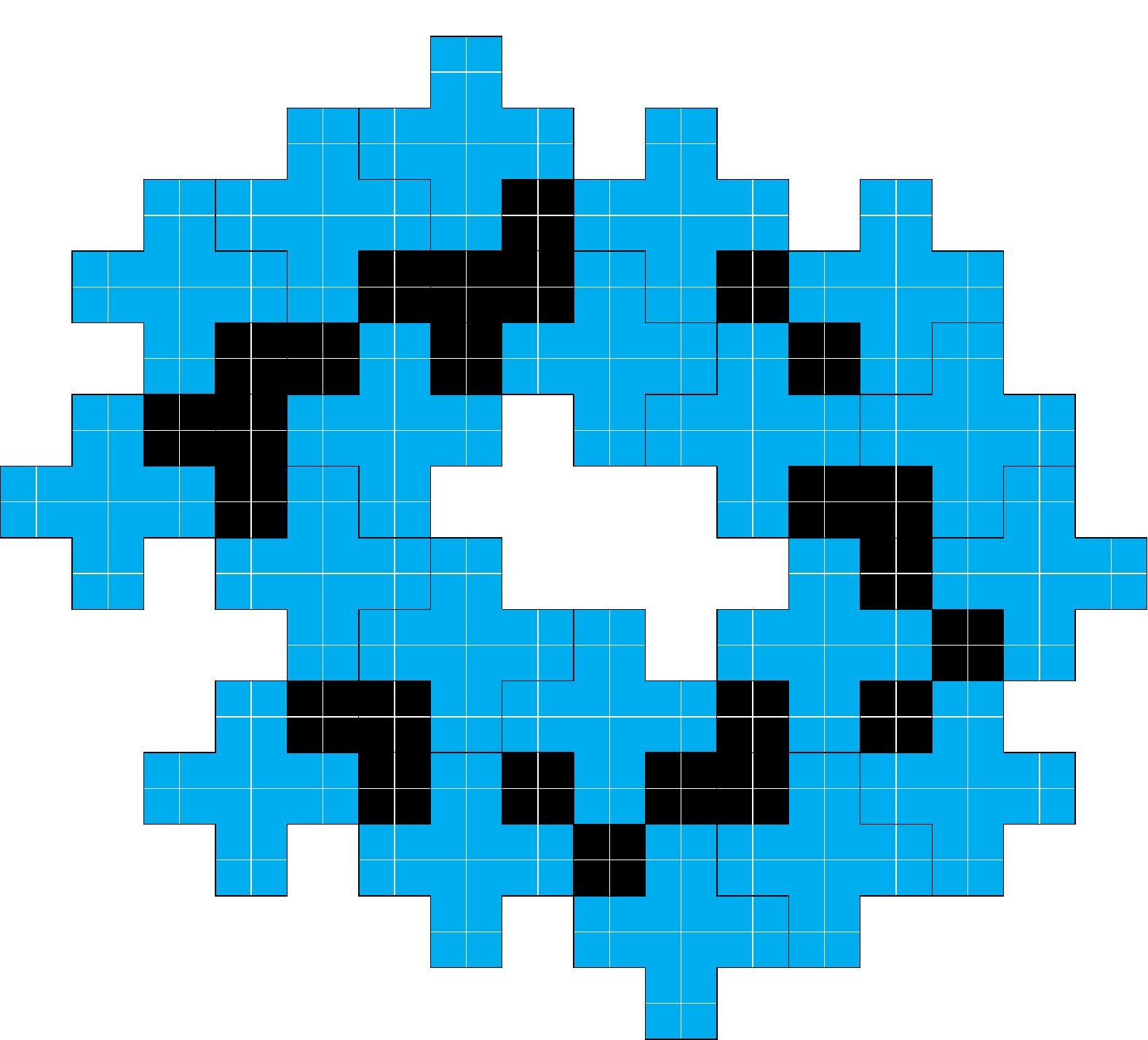}{\footnotesize\it c}.
  \caption{Example Gaunt-Fisher configurations. The empty space is colored black. Figure~\-{\it b} corresponds to the configuration in figure~\-\ref{fig:hole_example}{\it a}, and figure~\-{\it c} corresponds to that in figure~\-\ref{fig:interface}. The complement of these Gaunt-Fisher configurations can be entirely covered by crosses.}
  \label{fig:gaunt_fisher}
\end{figure}

\indent In the absence of sliding, it is easy to see which holes are correlated. Indeed, consider the connected components of the union of the holes and their neighboring particles (see figure~\-\ref{fig:gaunt_fisher}). Each such component is called a {\it Gaunt-Fisher (GF) configuration}. The space that is not occupied by GF configurations is completely covered by particles, so, since there is no sliding, each of its connected components can be extended to a perfect covering of $\Lambda$. To each such component, we associate an index that specifies to which of the possible perfect coverings the configuration in the component can be extended. The interaction between GF configurations is then rather simple: distinct GF configurations must be disjoint, and each connected component of the complement of each GF configuration must be coverable by particles in a configuration that can be extended to the perfect covering specified by the index of the component. The latter interaction is thus mediated by the indices of the connected components of the complements of the GF configurations. We then completely remove this interaction, as described below.
\bigskip

\begin{figure}
  \hfil
  \includegraphics[width=6cm]{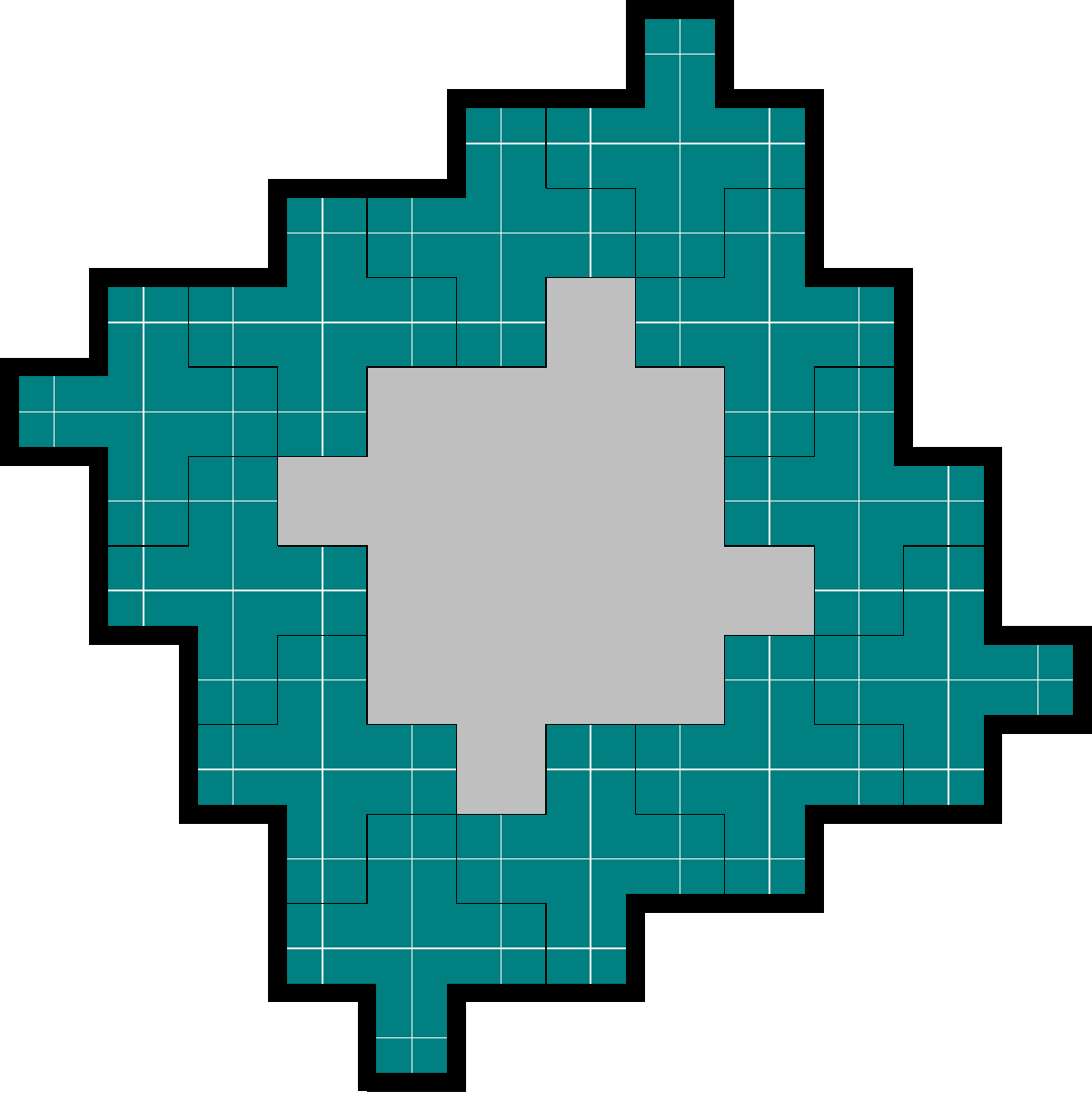}
  \hfil
  \includegraphics[width=6cm]{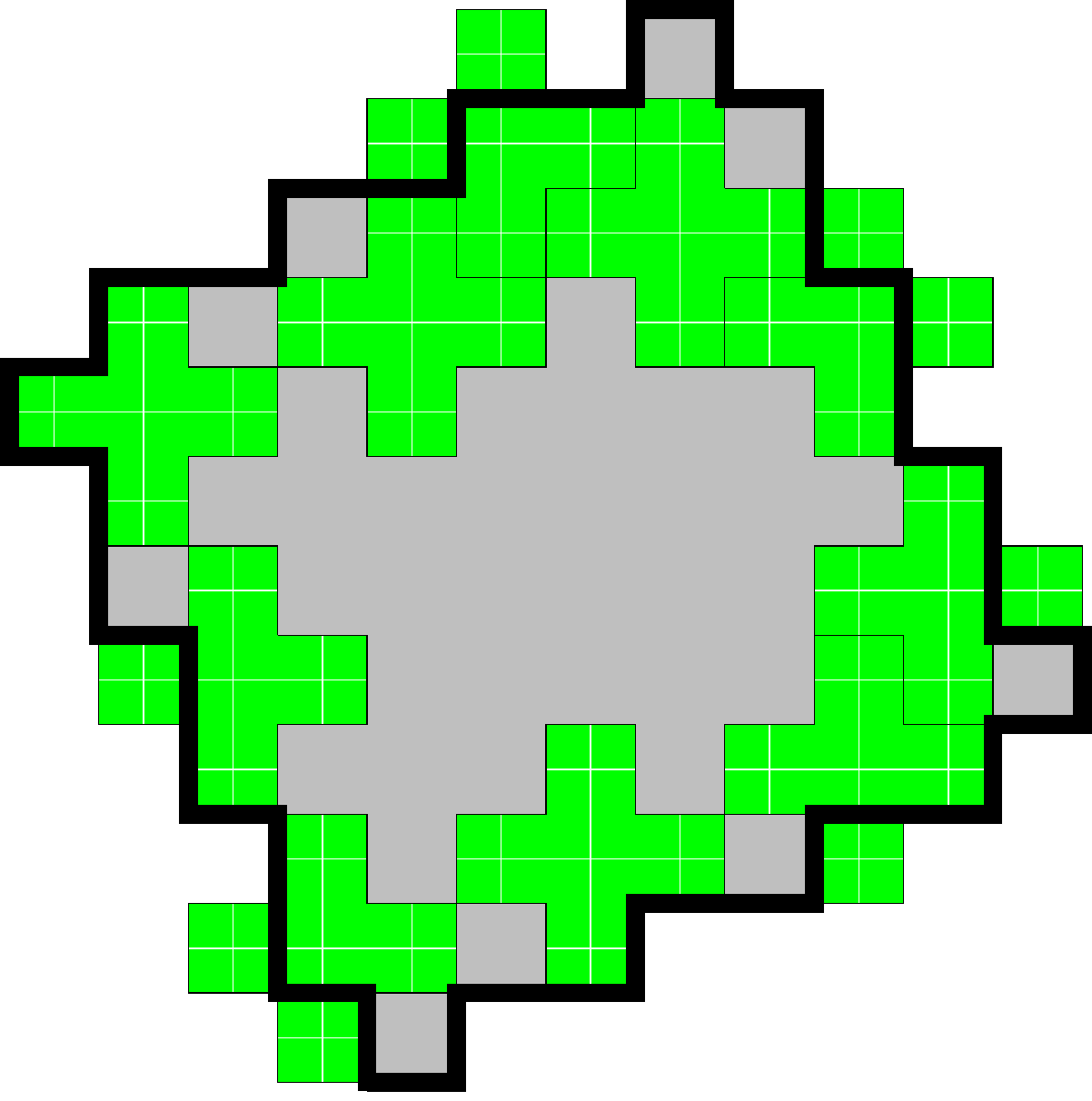}
  \caption{
    Two different boundary conditions for the hard cross model. The set $\Lambda$ is outlined by the thick black line. The crosses that are drawn are those mandated by the boundary condition (the boundary condition stipulates that every cross that is in contact with the boundary must be of a specified phase), and the remaining available space in $\Lambda$ is colored gray. The partition function in the case of figure~\-{\it a} is
    $$z^{16}(1+4y+10y^2+8y^3+y^4)$$
    whereas that in figure~\-{\it b} is
    $$z^{16}(1+6y+18y^2+48y^3+43y^4+13y^5+y^6).$$
  }
  \label{fig:assymmetry}
\end{figure}

\indent Let us consider only the most external GF configurations, that is those that do not lie inside any other GF configuration. To eliminate their interaction, we fix the boundary condition on $\Lambda$ once and for all, so as to fix the index of the perfect covering on the outside. At this point, the constraint on each GF configuration is that it be compatible with the boundary condition, which is independent of the other GF configurations, thus eliminating the interaction between them. Now, the GF configurations may have holes (see figure~\-\ref{fig:gaunt_fisher}{\it c}), and other configurations may lie inside these holes. These would interact with the GF configurations that contain them. In order to eliminate this interaction, we make use of a technique from Pirogov-Sinai theory~\-\cite{PS75,KP84}. Namely, the partition function in each hole is a partition function on a smaller volume, with the boundary condition imposed by the index of the covering of the hole. Now, in general, the partition function in a hole may depend on the boundary condition (see figure~\-\ref{fig:assymmetry}), but this dependence is weak. Indeed, one can show that the ratio of the partition function with one boundary condition divided by that with another is at most exponential in the size of the {\it boundary} (whereas each partition function is exponential in the size of the {\it bulk}). Thus, at the price of an exponential factor in the size of the GF configuration, called the {\it flipping term}, we can pretend that the partition function in each hole has the same boundary condition as the entire space. At this point, the interaction between GF configuration simply states that they must not overlap. And, since there is no sliding, each GF configuration contains a number of holes which is proportional to its size, thus contributing a factor $z^{-c|\mathrm{size}|}$ to the partition function (from which, we recall, we have factored out $z^{-\rho_m|\Lambda|}$ so that each hole contributes $z^{-\rho_m}$), which outweighs the flipping term.
\bigskip

\indent We have thus constructed a model of GF configurations, which interact via a purely hard-core potential, and which have a very small effective fugacity. We then use a low-fugacity expansion to express the function $f_\Lambda$ (\ref{f}) as a convergent series, following~\-\cite{KP86}. In addition, the effective fugacity of a GF configuration of volume $|V|$ is equal to the fugacity of the particles inside the configuration divided by $z^{\rho_m|V|}$. Furthermore, since the GF configuration contains empty space, the number of particles in the GF configuration is smaller than $\rho_m|V|$. Finally, since the volume obtained by removing the GF configuration from the lattice can be covered by particles, $\rho_m|V|$ is an integer. Therefore, the fugacity of a GF configuration is an analytic function of $y\equiv z^{-1}$, which implies that $f_\Lambda$ is analytic in $y$, uniformly in $|\Lambda|$.

\section{Concluding remarks}
\point In this paper, we have focused on the pressure $p$ at large $z$. Other thermodynamic quantities can be computed from $p$, or by a computation similar to that of $p$. We recover the average density $\bar\rho$ from the pressure by
\begin{equation}
  \bar\rho=z\frac{\partial p}{\partial z}=-y\frac{\partial p}{\partial y}=\rho_m+\sum_{j=1}^\infty kc_ky^k.
\end{equation}
Since the pressure is independent of the boundary condition, it is the same in all phases, which implies that the average density is as well. In order to distinguish between phases, one could consider the local density $\rho(x)$ at $x$, which does depend on the phase. Thus, for the diamonds on the square lattice, at large fugacities, the local density at sites on the even sublattice would be different from that on the odd sublattice: in the even phase, the former would be close to $1$, whereas the latter would be close to $0$. In general, when there are $n$ close-packed phases, there are $n$ sublattices, and, in each phase, the local density at one of the sublattices is close to $1$, while the others are close to $0$. The local density $\rho(x)$ can be expanded in powers of $y$, using similar methods to those described in this paper. Similarly, one can expand higher-order correlation functions, and find that, when the series converges, the truncated correlation functions in a specified phase decay exponentially.
\bigskip

\point Here, we have only considered HCLP systems that have a single shape. A natural extension would be to consider systems in which several types of particles of different shapes can coexist, provided there is a finite number of perfect coverings, and no sliding. In that case, different particles may have different fugacities, for instance, one might set $z_\alpha=\lambda_\alpha z$ and expand in $z^{-1}$. The qualitative behavior of the system might depend on the $\lambda_\alpha$. Further extensions could be to consider more general pair potentials, by, for instance, allowing for smooth interactions, or for more general hard-core repulsions, such as the Widom-Rowlinson~\-\cite{WR70} interaction.
\bigskip

\point If, instead of overlap between the particles being forbidden, it were merely discouraged by replacing the hard-core potential by a strongly repulsive potential $J$, then one would expect, using a technique similar to that sketched in this paper, to prove that the pressure is analytic in an intermediate regime $z_0<|z|\ll e^J$. This was shown for the soft diamond model in~\-\cite{BK73}.
\bigskip

\point The methods described here allow, in some cases, to approach the continuum, but not to reach it. For instance, in the cross model, one can make the lattice finer (or, equivalently, the crosses can be made thicker). However, the radius of convergence vanishes in the continuum limit. New ideas are needed to treat such a case.

\vfill
\hfil{\bf Acknowledgements}\par
\medskip
\indent We are grateful to Giovanni Gallavotti and Roman Koteck\'y for enlightening discussions. The work of J.L.L. was supported by AFOSR grant FA9550-16-1-0037. The work of I.J. was supported by The Giorgio and Elena Petronio Fellowship Fund and The Giorgio and Elena Petronio Fellowship Fund II.
\eject

\end{document}